\newif\iffigs\figstrue
\documentclass[12pt]{article}
\usepackage{latexsym,amssymb}
   \input{epsf}
\else
\fi

\newcommand{\sect}[1]{\setcounter{equation}{0}\section{#1}}

\newcommand{\eq}{\begin{equation}}
\newcommand{\eqa}{\begin{eqnarray}}
\newcommand{\en}{\end{equation}}
\newcommand{\ena}{\end{eqnarray}}
\newcommand{\enn}{\nonumber \end{equation}}

\def\sk{\vskip .4cm}
\def\noi{\noindent}
\def\om{\omega}

\def\al{\alpha}

\def\be{\beta}
\def\ga{\gamma}
\def\Ga{\Gamma}

\def\Cb{\bar{C}}

\def\rhop{{\rho}^{\prime}}

\def\we{\wedge}

\def\de{\delta}

\def\part{\partial}

\def\R#1#2{ R^{#1}_{~~~#2} }

\def\Cb{{\bf \mbox{\boldmath $C$}}}

\def\c#1#2{ C^{#1}_{~~#2} }
\def\C#1#2{ {\bf \mbox{\boldmath $C$}}^{#1}_{~~#2} }

\def\xp{x^{+}}

\def\gp{g^{\prime}}

\def\n2{{{N+1} \over 2}}

\def\square{{\,\lower0.9pt\vbox{\hrule \hbox{\vrule height 0.2 cm
\hskip 0.2 cm \vrule height 0.2 cm}\hrule}\,}}

\def\Q.E.D.{\rightline{$\Box$}}

\def\sumong{\sum_{g \in G}}
\def\sumongp{\sum_{g \in G'}}
\def\sumongnote{\sum_{g \not= e}}

\def\sumonh{\sum_{h \in G}}
\def\sumonhp{\sum_{h \in G'}}

\def\Lcal{{\cal L}}
\def\Rcal{{\cal R}}

\def\omc#1#2{\om^{#1}_{~~#2}}

\def\Gc#1#2{\Ga^{#1}_{~~#2}}
\def\a#1#2{a^{#1}_{~#2}}
\def\ainv#1#2{(a^{-1})^{#1}_{~#2}}

\def\xp{x'}
\def\gp{g'}

\def\Gcal{{\cal G}}
\def\unop{1  \! \mbox{I}}

\begin{document}
\begin{titlepage}
\vskip -1cm \rightline{DFTT-26/01}
\rightline{August 2001} \vskip 1em
\begin{center}
{\large\bf Finite group discretization of Yang-Mills and Einstein
actions }
\\[2em]
{\bf Leonardo Castellani}${}^{1,2}$ and {\bf Chiara
Pagani}${}^1$\\[.4em] {\sl ${}^1$ Dipartimento di Scienze e
Tecnologie Avanzate, Universit\`a del Piemonte Orientale, Italy;
\\${}^2$ Dipartimento di Fisica Teorica and I.N.F.N\\ Via P. Giuria 1,
10125 Torino, Italy.} \\ {\small castellani@to.infn.it}\\[2em]
\end{center}
\vskip 3 cm
\begin{abstract}
Discrete versions of the Yang-Mills and Einstein actions are
proposed for any finite group. These actions are invariant
respectively under local gauge transformations, and under the
analogues of Lorentz and general coordinate transformations. The
case $Z_n \times Z_n \times...\times Z_n$ is treated in some
detail, recovering the Wilson action for Yang-Mills theories, and
a new discretized action for gravity.
\end{abstract}

\vskip 8cm \noi \hrule \vskip .2cm
 \noi{\centerline {\footnotesize
  Supported in part by EC under RTN project
 HPRN-CT-2000-00131}}

\end{titlepage}
\newpage
\setcounter{page}{1}
\sect{Introduction}

Discretization of field theories, and in particular of gravity,
has a long history which we do not attempt to review here.
Comprehensive accounts and reference lists can be found in
\cite{revgravdiscrete}. The motivations to consider field theories
on discrete structures are at least of two kinds. One is
computational, as in quantum gauge theories, where the lattice
approach yields nonperturbative information. The other is related
to the mathematical consistency of the quantum theory, since
``separating" the spacetime points removes the ultraviolet
divergences: this is certainly useful in nonrenormalizable
theories like ordinary Einstein quantum gravity.

According to a current paradigm, quantum gravity arises within
string/brane theory \cite{stringbooks}, a consistent quantum
theory that is expected to unify all known interactions. Then
Einstein theory is to be considered a low energy effective field
theory, part of the more fundamental ``brane world" \footnote{In
fact in the brane world scenarios (see \cite{braneoldworld} for an
early reference, and \cite{RS} for a more contemporary point of
view) spacetime is considered as a dynamical brane within a higher
dimensional space.}, and as such not needing to make sense at the
quantum level by itself. But attempts to make it consistent are
still worthwhile: after all Yang-Mills theory, which also emerges
in the low energy regime of the string/brane theory, is a well
defined quantum field theory without need of a ``stringy"
regularization.

On the other hand, string/brane theory also suggests that
spacetime at short distances may not be smooth. For example gauge
theories on noncommutative spaces are a low-energy limit of open
strings in a background $B$-field, describing the fluctuations of
the D-brane worldvolume \cite{stringynoncommgauge,seibergwitten}.
This has prompted many investigations in $*$ - deformations of
Yang-Mills theories \footnote{ In contrast, there are very few
studies for $*$ - deformed gravity theories, the reason being that
one lacks a ``stringy" motivation. But it would be of interest to
see how the intrinsic nonpolynomiality of gravity plays against
the nonpolynomiality of the $*$ product.}, and has provided one of
the bridges between string theory and noncommutative geometry
(NCG). Reviews on  NCG can be found in \cite{NCGreviews}.

In this paper we construct a discretized Yang-Mills and Einstein
action for any finite group $G$: this is a particular
noncommutative setting, where noncommutativity does not concern
coordinates between themselves, but only coordinates with
differentials.

The spacetime points are replaced by isolated points labeling the
group elements of $G$. The functions on these sets of points are
endowed with differential calculi, due to the Hopf algebra induced
by the group structure \cite{DMGcalculus,FI1,gravfg,tmr} Then one
can construct the analogue of the vielbein, the connection, the
curvature etc. on these finite group ``manifolds". In fact this
program can be carried out for any Hopf algebra, quantum groups
being a notable example (differential calculi on Hopf algebras
were first constructed in \cite{wor}; for a review with
applications to field theory see for ex. \cite{ACintro}). Here we
use the differential calculi to define Yang-Mills and Einstein
actions
 on any finite $G$, and we show that these actions
 are respectively invariant under the $G$-analogues of local
gauge variations, and of Lorentz and general coordinate
transformations. Previous investigations on finite $G$ field
theories can be found in \cite{DMgauge,DMgrav,gravfg,tmr,majid}.

A possible application of this technique is to use finite group
spaces as internal spaces in Kaluza-Klein (super)gravity or
superstring theories. Harmonic analysis on finite group spaces
being elementary, the reduction of the higher dimensional theory
is easy to implement. In fact the Kaluza-Klein reduction of $M^4
\times Z_2$ gauge theories coupled to fermions yields a Higgs
field in $d=4$, with the correct spontaneous symmetry breaking
potential and Yukawa couplings, see for ex. \cite{tmr}.

In Section 2 we give a review of the differential calculus on
finite groups. Section 3 illustrates the general results for the
case $G=(Z_n)^N$. In Section 4 we present the Yang-Mills action on
finite groups. Section 5 contains two proposals for a gravity
action on a finite group $G$, differing in the choice of tangent
group, and the corresponding invariances are discussed; the
$G=(Z_n)^N$ gravity action is treated in more detail. Some
conclusions are included in Section 6.

\section{Differential calculus on finite groups}

\sk
   \noi {\bf Notations}
    \sk
    Let $G$ be a finite group of order $n$ with
generic element $g$ and unit $e$. Consider $Fun(G)$, the set of
complex functions on $G$. An element $f$ of $Fun(G)$ is specified
by its values $f_g \equiv f(g)$ on the group elements $g$, and can
be written as
 \eq
  f=\sum_{g \in G} f_g x^g,~~~f_g \in \Cb \label{fonG}
 \en
where the functions $x^g$ are defined by
 \eq
  x^g(g') = \de^g_{g'} \label{xg}
\en
Thus $Fun(G)$ is a n-dimensional vector space, and the $n$
functions $x^g$ provide a basis. $Fun(G)$ is also a commutative
algebra, with the usual pointwise sum and product, and unit $I$
defined by $I(g)=1, \forall g \in G$. In particular: \eq x^g
x^{g'}=\de_{g,g'} x^g,~~~\sumong x^g = I \label{mul}
\en
The left action of the group on itself induces the (pullback)
$\Lcal_{g_1}$ on $Fun(G)$: \eq \Lcal_{g_1} f(g_2) \equiv
f(g_1g_2)|_{g_2},~~~\Lcal_{g_1}:Fun(G) \rightarrow Fun(G)
\en
where $f(g_1g_2)|_{g_2}$ means $f(g_1g_2)$ seen as a function of
$g_2$. Similarly we can define the right action  on $Fun(G)$ as:
 \eq
(\Rcal_{g_1}f)(g_2)= f(g_2g_1)|_{g_2}
\en
For the basis functions we find easily: \eq \Lcal_{g_1} x^{g} =
x^{g_1^{-1} g}, ~~\Rcal_{g_1} x^{g} = x^{g g_1^{-1}}
\en
Moreover:
 \eqa & &\Lcal_{g_1} \Lcal_{g_2}=\Lcal_{g_2g_1},
~~\Rcal_{g_1} \Rcal_{g_2}=\Rcal_{g_1g_2},\\ & &\Lcal_{g_1}
\Rcal_{g_2}=\Rcal_{g_2} \Lcal_{g_1}
 \ena
 The $G$ group structure induces a Hopf algebra structure on
$Fun(G)$, and this allows the construction of differential calculi
on $Fun(G)$, according to the techniques of ref.
\cite{wor,ACintro}. We summarize here the main properties of these
calculi. A detailed treatment can be found in \cite{gravfg}.

  \sk
   \noi {\bf Exterior differential}
    \sk

A (first-order) differential calculus on $Fun(G)$ is defined by a
linear map $d$: $Fun(G) \rightarrow \Gamma$, satisfying the
Leibniz rule \eq
 d(ab)=(da)b+a(db),~~\forall a,b\in Fun(G);
\label{Leibniz}
\en
The ``space of 1-forms" $\Ga$ is an appropriate bimodule on
$Fun(G)$, which essentially means that its elements can be
multiplied on the left and on the right by elements of $Fun(G)$.
{} From the Leibniz rule $da=d(Ia)=(dI)a+Ida$ we deduce $dI=0$.
Consider the differentials of the basis functions $x^g$. From
$0=dI=d(\sumong x^g)=\sumong dx^g$ we see that only $n-1$
differentials are independent. Moreover every element of $\Ga$ can
be expressed as a linear combination (with complex coefficients)
of terms of the type $x^g dx^{g'}$, since the commutations: \eq
dx^g x^{g'} = -x^g dx^{g'}+\de^g_{g'} dx^g
\en
allow to reorder functions to the left of differentials.
 \sk
 \noi {\bf Partial derivatives}
 \sk
  Consider the differential of a
function $f \in Fun(g)$: \eq df = \sumong f_g dx^g = \sumongnote
f_g dx^g + f_e dx^e= \sumongnote (f_g - f_e)dx^g \equiv
\sumongnote
\part_g f dx^g \label{partcurved}
\en
We have used $dx^e = - \sumongnote dx^g$ (from $\sumong dx^g=0$).
The partial derivatives of $f$ have been defined in analogy with
the usual differential calculus, and are given by \eq
\part_g f = f_g - f_e = f(g) - f(e) \label{partcurved2}
\en
Not unexpectedly, they take here the form of finite differences
(discrete partial derivatives at the origin $e$). \sk

 \noi
 {\bf Left and right covariance}
\sk A differential calculus is
 {\sl left or right covariant} if the left or right action of
 $G$ ($\Lcal_g$ or $\Rcal_g$) commutes with the exterior derivative $d$.
 Requiring left and right covariance in fact {\sl defines} the action of
 $\Lcal_g$ and $\Rcal_g$ on differentials: $\Lcal_g db \equiv
 d(\Lcal_g b), \forall b \in Fun(G)$ and similarly for
 $\Rcal_g db$. More generally, on elements of $\Ga$
 (one-forms) we define $\Lcal_g$ as:
 \eq
 \Lcal_g (adb) \equiv (\Lcal_g a) \Lcal_g db =
 (\Lcal_g a) d (\Lcal_g b)
 \en
 and similar for $\Rcal_g$.
 A differential calculus is called {\sl bicovariant} if it is
both left and right covariant.

\sk \noi {\bf Left and right invariant one forms}
 \sk
 As in usual Lie group manifolds, we can introduce a basis in $\Ga$
 of left-invariant one-forms $\theta^g$:
 \eq
  \theta^g \equiv
\sumonh x^{hg} dx^h ~~(=\sumonh x^h dx^{hg^{-1}}),
\label{deftheta}
\en
It is immediate to check that indeed $\Lcal_k \theta^g =
\theta^g$. The relations (\ref{deftheta}) can be inverted:
 \eq
  dx^h = \sumong (x^{hg} - x^h)\theta^g \label{dxastheta}
\en
{} From $0=dI=d\sumong x^g =\sumong dx^g=0$ one finds: \eq
 \sumong
\theta^g = \sum_{g,h \in G} x^h dx^{hg^{-1}}= \sumonh x^h \sumong
dx^{hg^{-1}}=0 \label{sumtheta}
\en
Therefore we can take as basis of the cotangent space $\Ga$ the
$n-1$ linearly independent left-invariant one-forms $\theta^g$
with $g \not= e$ (but smaller sets of $\theta^g$ can be
consistently chosen as basis, see later).

 Analogous results hold
for right invariant one-forms $\zeta^g$:
 \eq
  \zeta^g = \sumonh x^{gh}dx^h
\en
{} From the expressions of $\theta^g$ and $\zeta^g$ in terms of
$xdx$, one finds the relations
 \eq
  \theta^g = \sumonh x^h \zeta^{ad(h)g},~~~
  \zeta^g=\sumonh x^h \theta^{ad(h^{-1})g}
\en
 \noi {\bf Conjugation}
 \sk
 On $Fun(G)$ there are two natural involutions $*$ satisfying $(ab)^*=b^*a^*$
 ($=a^*b^*$ since functions on $G$ commute):
 \eqa
 & & (x^g)^*=x^g   \label{conj1}\\
 & & (x^g)^\star = x^{g^{-1}} \label{conj2}
 \ena
 We use the slightly different symbol $\star$ for the second one.
 These conjugations are extended to the (first-order) differential
 calculus via the rule:
 \eq
 (x^h dx^k)^*=(dx^k)^* (x^h)^*
 \en
 and similar for $\star$. Then
 \eq
 (\theta^g)^*=-\theta^{g^{-1}},~~~(\theta^g)^\star=-\zeta^g
 \en
\noi {\bf Commutations between $x$ and $\theta$ }
 \eq
  x^h dx^g = x^h \theta^{g^{-1}h} =
\theta^{g^{-1}h} x^g ~~(h\not=g)~~\Rightarrow x^h \theta^g =
\theta^g x^{hg^{-1}}~~(g\not=e) \label{xthetacomm}
\en
implying the general commutation rule between functions and
left-invariant one-forms: \eq f \theta^g = \theta^g \Rcal_g f
\label{fthetacomm}
\en
Thus functions do commute between themselves (i.e. $Fun(G)$ is a
commutative algebra) but do not commute with the basis of
one-forms $\theta^g$. In this sense the differential geometry of
$Fun(G)$ is noncommutative. \sk
 \noi
 {\bf Classification of bicovariant calculi}
 \sk
 The right action of $G$ on the elements
$\theta^g$ is given by: \eq \Rcal_h \theta^g =
\theta^{ad(h)g},~~\forall h \in G
\en
where $ad$ is the adjoint action of $G$ on itself, i.e. $ad(h)g
\equiv hgh^{-1}$. Then {\sl bicovariant calculi are in 1-1
correspondence with unions of conjugacy classes (different from
$\{e\}$)} \cite{DMGcalculus}: if $\theta^g$ is set to zero, one
must set to zero all the $\theta^{ad(h)g},~\forall h \in G$
corresponding to the whole conjugation class of $g$.
 \sk
 We denote by $G'$ the subset corresponding
 to  the union of conjugacy classes
 that characterizes the bicovariant calculus on $G$
 ($G' = \{g \in G |\theta^g \not= 0\}$).
 Unless otherwise indicated, repeated indices are
 summed on $G'$ in the following.
 \sk

 \noi {\bf Exterior product}
\sk An {\sl exterior product}, compatible with the left and right
actions of $G$, can be defined by \eq \theta^{g_1} \we
\theta^{g_2}=\theta^{g_1} \otimes \theta^{g_2} - \theta^{g_1^{-1}
g_2 g_1} \otimes \theta^{g_1}
 \label{exprod}
\en
where the tensor product between elements $\rho,\rhop \in \Ga$ is
defined to have the properties $\rho a\otimes \rhop=\rho \otimes a
\rhop$, $a(\rho \otimes \rhop)=(a\rho) \otimes \rhop$ and $(\rho
\otimes \rhop)a=\rho \otimes (\rhop a)$.

Note that:
 \eq
  \theta^{g} \we \theta^{g}=0~~~~\mbox{(no sum on $g$)}
\en

  Left and right actions on $\Ga \otimes \Ga$ are
  simply defined by:
  \eqa
 & & \Lcal_h (\rho \otimes \rhop)= \Lcal_h \rho \otimes \Lcal_h
  \rhop,~~~\\
& &\Rcal_h (\rho \otimes \rhop)= \Rcal_h \rho \otimes \Rcal_h
  \rhop
  \ena
 Compatibility  of the exterior product with $\Lcal$ and $\Rcal$
 means that
 \eq
 \Lcal(\theta^i \we \theta^j)=\Lcal\theta^i \we \Lcal
 \theta^j, ~~\Rcal(\theta^i \we \theta^j)=\Rcal\theta^i \we \Rcal
 \theta^j
 \en
 Only the second relation is nontrivial and is verified upon use
 of the definition (\ref{exprod}).

  The generalization  to exterior products of $n$
one-forms is straightforward, see ref. \cite{gravfg}
 \sk
 \noi {\bf Exterior derivative}
 \sk
  Equipped with the exterior product we can
define the {\sl exterior derivative} \eq d~:~\Gamma \rightarrow
\Gamma \we \Gamma
\en
\eq d (a_k db_k) = da_k \we db_k,
\en
\noi which can easily be extended to $\Gamma^{\we n}$ ($d:
\Gamma^{\we n} \rightarrow \Gamma^{\we (n+1)}$), and has the
following properties: \eq
 d(\rho \we \rhop)=d\rho \we \rhop +
(-1)^k \rho \we d\rhop \label{propd1}
\en
\eq d(d\rho)=0\label{propd2}
\en
\eq \Lcal_g (d\rho)=d \Lcal_g \rho \label{propd3}
\en
\eq \Rcal_g (d\rho)=d \Rcal_g \rho \label{propd4}
\en
\noi where $\rho \in \Ga^{\we k}$, $\rhop \in \Ga^{\we n}$,
$\Ga^{\we 0} \equiv Fun(G)$ . The last two properties express the
fact that $d$ commutes with the left and right action of $G$. \sk
\noi {\bf Tangent vectors} \sk Using (\ref{dxastheta}) to expand
$df$ on the basis of the left-invariant one-forms $\theta^g$
defines the (left-invariant) tangent vectors $t_g$:
 \eqa
 & & df=\sumong f_g dx^g  = \sumonhp (\Rcal_{h^{-1}} f - f ) \theta^h
\equiv \nonumber \\ & &~~~\equiv \sumonhp (t_h f) \theta^h
\label{partflat} \ena so that the ``flat" partial derivatives $t_h
f$ are given by \eq t_h f = \Rcal_{h^{-1}} f - f \label{partflat2}
\en
 The Leibniz rule for the flat partial derivatives $t_g$
reads: \eq
 t_g (ff')=
 (t_g f)  f'  +\Rcal_{g^{-1}}(f) t_g f'
 =(t_g f) \Rcal_{g^{-1}} f'  + f t_g f' \label{tgLeibniz}
\en

In analogy with ordinary differential calculus, the operators
$t_g$ appearing in (\ref{partflat}) are called (left-invariant)
{\sl tangent vectors}, and in our case are given by
 \eq
  t_g =
\Rcal_{g^{-1}}- id \label{tangent}
\en
\noi {\bf Fusion algebra}
 \sk The tangent vectors satisfy the fusion algebra:
  \eq
t_g t_{g'}= \sum_h \c{h}{g,g'} t_h \label{chichi}
\en
where the structure constants are: \eq \c{h}{g,g'}=\de^h_{g'g} -
\de^h_{g}-\de^h_{g'} \label{cconst}
\en
and are $ad(G)$ invariant: \eq \c{ad(h)g_1}{~~ad(h)g_2,ad(h)g_3}=
\c{g_1}{g_2,g_3} \label{adhc}
\en
Defining:
 \eq
\C{g}{g_1,g_2} \equiv \c{g}{g_1,g_2} - \c{g}{g_2,g_2 g_1
g_2^{-1}}=\de^{ad(g_2^{-1})g}_{g_1} - \de^g_{g_1} \label{Cconst}
\en
the following fusion identities hold:
 \eq \C{k}{h_1,g} \C{h_2}{k,g'}=
\c{h}{g,g'} \C{h_2}{h_1,h} \label{adfusion}
\en

Thus the $\Cb$ structure constants are a representation (the
adjoint representation) of the tangent vectors $t$. Besides
property (\ref{adhc}) they also satisfy:
 \eq
\C{g}{g_1,g_2}=\C{g_1}{g,g_2^{-1}} \label{propC}
\en
 \sk
  \noi
   {\bf Cartan-Maurer equations, connection and curvature}
 \sk
  From the
definition (\ref{deftheta}) and eq. (\ref{fthetacomm}) we deduce
the Cartan-Maurer equations:
 \eq
 d\theta^g + \sum_{g_1,g_2}
\c{g}{g_1,g_2}\theta^{g_1}\we \theta^{g_2}=0 \label{CM}
\en
where the structure constants $\c{g}{g_1,g_2}$ are those given in
(\ref{cconst}). \sk Parallel transport of the vielbein $\theta^g$
can be defined as in ordinary Lie group manifolds: \eq \nabla
\theta^g= - \omc{g}{g'} \otimes \theta^{g'}
 \label{parallel}
\en
where $\omc{g_1}{g_2}$ is the connection one-form: \eq
\omc{g_1}{g_2}= \Gc{g_1}{g_3,g_2} \theta^{g_3}
\en
Thus parallel transport is a map from $\Ga$ to $\Ga \otimes \Ga$;
by definition it must satisfy: \eq \nabla (a \rho) = (da)\otimes
\rho + a \nabla \rho,~~~\forall a \in A,~\rho \in \Ga
\label{parallel1}
\en
and it is a simple matter to verify that this relation is
satisfied with the usual parallel transport of Riemannian
manifolds. As for the exterior differential, $\nabla$ can be
extended to a map $\nabla : \Ga^{\we n} \otimes \Ga
\longrightarrow \Ga^{\we (n+1)} \otimes \Ga $ by defining:
 \eq
\nabla (\varphi \otimes \rho)=d\varphi \otimes \rho +
 (-1)^n \varphi \nabla
\rho
\en

Requiring parallel transport to commute with the left and right
action of $G$ means:
 \eqa & &\Lcal_{h} (\nabla \theta^{g})=\nabla
( \Lcal_{h} \theta^{g}) =\nabla \theta^g\\ & &\Rcal_{h} (\nabla
\theta^{g})=\nabla ( \Rcal_{h} \theta^{g}) =\nabla \theta^{ad(h)g}
\ena
 Recalling that  $\Lcal_{h} (a \rho)=(\Lcal_h a) (\Lcal_h
\rho)$ and $\Lcal_{h} (\rho \otimes \rho')=(\Lcal_h \rho) \otimes
(\Lcal_h \rho'),~\forall a \in A,~\rho,~\rho' \in \Ga$ (and
similar for $\Rcal_h$),
 and substituting
(\ref{parallel}) yields respectively: \eq \Gc{g_1}{g_3,g_2} \in
\Cb
\en
and
 \eq
 \Gc{ad(h)g_1}{ad(h)g_3,ad(h)g_2}=\Gc{g_1}{g_3,g_2} \label{adga}
\en
Therefore the same situation arises as in the case of Lie groups,
for which  parallel transport on the group manifold commutes with
left and right action iff the connection components are $ad(G)$ -
conserved constant tensors. As for Lie groups, condition
(\ref{adga}) is satisfied if one takes $\Ga$ proportional to the
structure constants. In our case, we can take any combination of
the $C$ or $\Cb$ structure constants, since both are $ad(G)$
conserved constant tensors. As we see below, the $C$ constants can
be used to define a torsionless connection, while the $\Cb$
constants define a parallelizing connection.

\sk
 As usual, the {\sl curvature} arises from $\nabla^2$:
  \eq
  \nabla^2 \theta^g = - \R{g}{g'} \otimes \theta^{g'}
\en
\eq \R{g_1}{g_2} \equiv d \omc{g_1}{g_2} + \omc{g_1}{g_3} \we
\omc{g_3}{g_2} \label{curvature}
\en

The {\sl torsion} $R^g$ is defined by: \eq R^{g_1} \equiv
d\theta^{g_1} +  \omc{g_1}{g_2} \we \theta^{g_2} \label{torsion}
\en

Using the expression of $\om$ in terms of $\Ga$ and the
Cartan-Maurer equations yields
 \eqa
& & \R{g_1}{g_2} =
 (- \Gc{g_1}{h,g_2}
\c{h}{g_3,g_4} + \Gc{g_1}{g_3,h} \Gc{h}{g_4,g_2})~ \theta^{g_3}
\we \theta^{g_4}=\nonumber \\ & & = (- \Gc{g_1}{h,g_2}
\C{h}{g_3,g_4} + \Gc{g_1}{g_3,h} \Gc{h}{g_4,g_2}- \Gc{g_1}{g_4,h}
\Gc{h}{g_4g_3g_4^{-1},g_2})~\theta^{g_3} \otimes
 \theta^{g_4}\nonumber
\ena \eqa & & R^{g_1}= (- \c{g_1}{g_2,g_3} + \Gc{g_1}{g_2,g_3})~
\theta^{g_2} \we \theta^{g_3}= \nonumber \\ & & (-
\C{g_1}{g_2,g_3} + \Gc{g_1}{g_2,g_3}-
\Gc{g_1}{g_3,g_3g_2g_3^{-1}})\theta^{g_2} \otimes \theta^{g_3}
\ena

Thus a connection satisfying:
 \eq
  \Gc{g_1}{g_2,g_3}-
\Gc{g_1}{g_3,g_3g_2g_3^{-1}}=\C{g_1}{g_2,g_3} \label{rconn}
  \en
corresponds to a vanishing torsion $R^g =0$ and could be
  referred to as a ``Riemannian" connection.
\sk
 On the other hand,  the choice:
   \eq
  \Gc{g_1}{g_2,g_3}=\C{g_1}{g_3,g_2^{-1}} \label{parconn}
  \en
corresponds to a vanishing curvature $\R{g}{g'}=0$, as can be
checked by using the fusion equations (\ref{adfusion}) and
property (\ref{propC}). Then (\ref{parconn}) can be called the
parallelizing connection: {\sl finite groups are parallelizable.}
\sk
 \noi {\bf Tensor transformations } \sk Under the
familiar transformation of the connection 1-form: \eq
(\omc{i}{j})' = \a{i}{k} \omc{k}{l} \ainv{l}{j} + \a{i}{k} d
\ainv{k}{j} \label{omtransf}
\en
the curvature 2-form transforms homogeneously: \eq (\R{i}{j})' =
\a{i}{k} \R{k}{l} \ainv{l}{j} \label{Rtransf}
\en

 \sk
  \noi
   {\bf Metric}
\sk
 The metric tensor $\ga$ can be defined as an element of $\Ga
\otimes \Ga$:
 \eq
  \ga = \ga_{i,j} \theta^i \otimes \theta^j
  \en
 Requiring it to be invariant under left and right action of
 $G$ means:
 \eq
 \Lcal_h (\ga)=\ga=\Rcal_h (\ga)
 \en
or equivalently, recalling $\Lcal_h(\theta^i \otimes
\theta^j)=\theta^i \otimes \theta^j$, $\Rcal_h(\theta^i \otimes
\theta^j)=\theta^{ad(h)i}\otimes \theta^{ad(h)j}$  :
 \eq
  \ga_{i,j} \in \Cb,~~  \ga_{ad(h)i,ad(h)j}=\ga_{i,j} \label{gabiinv}
 \en
These properties are analogous to the ones satisfied by the
Killing metric of Lie groups, which is indeed constant and
invariant under the adjoint action of the Lie group.
 \sk
 On finite $G$ there are various choices of biinvariant
 metrics. One can simply take $\ga_{i,j}=\de_{i,j}$,
   or $\ga_{i,j}= \C{k}{l,i} \C{l}{k,j}$.
   \sk
 For any biinvariant metric $\ga_{ij}$ there are tensor
 transformations (isometries)
 $\a{i}{j}$ under which $\ga_{ij}$ is invariant, i.e.:
 \eq
 \a{h}{h'} \ga_{h,k} \a{k}{k'}=\ga_{h',k'} \Leftrightarrow
\a{h}{h'} \ga_{h,k} = \ga_{h',k'} \ainv{k'}{k} \label{ginv}
 \en
 A class of isometries has been discussed in ref. \cite{gravfg}.
 In the case $\ga_{i,j}=\de_{i,j}$ the isometries are clearly given by
 the usual orthogonal matrices.

\sk
 \noi
  {\bf Lie derivative and diffeomorphisms}
   \sk
   The analogue of infinitesimal diffeomorphisms is
found using general results valid for Hopf algebras
\cite{ACintro,LCqgravity,LCqgravitycmp,Athesis}, of which finite
groups are a simple example. As for differentiable manifolds, it
can be expressed via the Lie derivative, which for finite groups
takes the form:
 \eq
 l_{t_g} \rho =  [\Rcal_{g^{-1}} \rho - \rho] \label{Liederivative2}
 \en
where $\rho$ is an arbitrary form field. Thus the Lie derivative
along $t_g$ coincides with the tangent vector $t_g$.

 As in the case of differentiable manifolds, the Cartan formula
 for the Lie derivative acting on $p$-forms holds:
\eq
 l_{t_g}= i_{t_g} d + d i_{t_g}
 \en
see ref.s
\cite{ACintro,AS,LCqgravity,LCqgravitycmp,Athesis,gravfg}.

 Exploiting this formula, diffeomorphisms
  (Lie derivatives) along generic tangent vectors $V$
 can also be consistently defined via the operator:
\eq
 l_{V}= i_{V} d + d i_{V}
 \en
 This requires
  a suitable definition
 of the contraction operator $i_V$  along generic tangent vectors
 $V$, discussed in ref.s \cite{AS,LCqgravitycmp,Athesis,gravfg}.

 We have then a way
 of defining ``diffeomorphisms" along arbitrary (and x-dependent)
 tangent vectors for any tensor $\rho$:
 \eq
 \delta \rho = l_V \rho
 \en
and of testing the invariance of candidate lagrangians under the
generalized Lie derivative.
 \sk
 \noi {\bf Finite coordinate transformations}
 \sk
The basis functions $x^g$ defined in (\ref{xg}) are the
``coordinates" of $Fun(G)$. The most general coordinate
transformation takes the form:
 \eq
 \xp^{\gp} = \sum_{g \in G} \xp^{\gp}_g x^g \label{gct}
 \en
where the $n \times n$  matrix $\xp^{\gp}_g \in GL(n,\mathbb{C})$
is a constant invertible matrix. An example will be provided in
the $Z^N \times Z^N \times...\times Z^N$ case. Let's consider now
the transformation of the differentials:
 \eq
 d\xp^{\gp} = \sumong  \xp^{\gp}_g dx^g = \sumongnote \xp^{\gp}_g
 dx^g + \xp^{\gp}_e dx^e = \sumongnote (\xp^{\gp}_g -\xp^{\gp}_e)
 dx^g = \sumongnote \part_g \xp^{\gp} dx^g
 \en
 a formula quite similar to the usual one, the only subtlety
being that the index $g$ does not include $e$. The $(n-1) \times
(n-1)$ constant matrix $A^{\gp}_{~~g} \equiv \part_g \xp^{\gp}$
belongs then to $ GL(n-1,\mathbb{C})$. This holds for the
universal calculus, with $n-1$ independent differentials. Then the
$p$-forms $dx^{g_1} \we...dx^{g_p}$ transform as covariant
tensors. When the independent $dx^g$ (or equivalently the
independent $\theta^g$) are less than $n-1$, the matrix
$A^{\gp}_{~~g}$ is not constant any more, as we'll see in the case
of $Z^N$, and exterior products of coordinate differentials do not
transform covariantly any more, due to noncommutativity of
$A^{\gp}_{~~g}(x)$ with $dx^g$.

 It is easy to prove the formula for the transformation of
the partial derivatives:
 \eq
 \part_{\gp} = \sumongnote (A^{-1})^g_{~~\gp} \part_g
 \en

 \sk
  \noi {\bf Haar measure and integration}
   \sk
    Since we want to define actions (integrals on
$p$-forms), we must now define integration of $p$-forms on finite
groups.

 Let us start with integration of functions $f$. We define the integral
  map $h$ as a linear functional $h: Fun(G) \mapsto \mathbb{C}$ satisfying the
  left and right invariance conditions:
  \eq
  h(\Lcal_g f)=h(f)=h(\Rcal_g f)
  \en
  Then this map is uniquely determined (up to a normalization constant),
  and is simply given by the ``sum over $G$" rule:
  \eq
  h(f)= \sumong f(g)
  \en

 Next we turn to define the integral of a $p$-form.
Within the differential calculus we have a basis of left-invariant
1-forms, which allows the definition of a biinvariant volume
element. In general for a differential calculus with $m$
independent tangent vectors, there is an integer $p  \geq m$ such
that the linear space of $p$-forms is 1-dimensional, and $(p+1)$-
forms vanish identically \footnote{with the exception of $Z_2$,
see ref. \cite{tmr}}. This means that every product of $p$ basis
one-forms $\theta^{g_1} \we \theta^{g_2} \we ... \we \theta^{g_p}$
is proportional to one of these products, that can be chosen to
define the volume form $vol$:
 \eq
 \theta^{g_1} \we \theta^{g_2} \we ... \we \theta^{g_p}=
 \epsilon^{g_1,g_2,...g_p} vol \label{vol}
 \en
 where $\epsilon^{g_1,g_2,...g_p}$ is the proportionality constant.
 The volume $p$-form is obviously left invariant. As shown in ref. \cite{gravfg}
 it is also right invariant, and the proof is based on the $ad(G)$ invariance of
 the $\epsilon$ tensor:
 $\epsilon^{ad(g)h_1,...ad(g)h_p}=\epsilon^{h_1,...h_p}$.

Having identified the volume $p$-form it is natural to set
 \eq
 \int f vol \equiv h(f) = \sumong f(g) \label{intpform}
 \en
 and  define the integral on a $p$-form $\rho$ as:
 \eqa
 & & \int \rho = \int \rho_{g_1,...g_p}~ \theta^{g_1}
 \we ... \we \theta^{g_p}=  \nonumber \\
 & & \int
\rho_{g_1,...g_p}~\epsilon^{g_1,...g_p} vol \equiv \nonumber \\ &
& \equiv ~~~ \sumong \rho_{g_1,...g_p}(g)~\epsilon^{g_1,...g_p}
  \ena
Due to the biinvariance of the volume form, the integral map $\int
: \Ga^{\we p} \mapsto \Cb$ satisfies the biinvariance conditions:
 \eq
  \int \Lcal_g f = \int f = \int \Rcal_g f
  \en

  Moreover, under the assumption that
  $d(\theta^{g_2} \we ... \we \theta^{g_p})=0$, i.e.
  that any exterior product of $p-1$ left-invariant one-forms $\theta$ is closed,
 the important property holds:
  \eq
  \int df =0
  \en
  with $f$  any $(p-1)$-form: $f=f_{g_2,...g_p}~ \theta^{g_2}
\we ... \we \theta^{g_p}$. This property, which allows
  integration by parts, has a simple proof (see ref. \cite{gravfg}).

\section{Calculus on $Z_n \times ...\times Z_n$}

We apply here the general theory to products of cyclic groups. For
simplicity we assume the order of these cyclic groups to be the
same.

We start with $Z_n$ and then generalize to products.
 \sk
 \noi {\bf Calculus on $Z_n$}

 \sk
 \noi {\sl Elements of $Z_n$}: $u^i = \{e,u,u^2,...u^{n-1}\}$, with
$u^0=u^n=e$.
 \sk
 \noi {\sl Basis of dual functions on $Z_n$}:
 $x^{u^i} = \{x^e,x^u,x^{u^2},...,x^{u^{n-1}} \}$. Left and right actions
 coincide since the group is abelian, i.e. $\Lcal_{u^{j}} x^{u^i}=
 x^{u^{i-j}} = \Rcal_{u^{j}} x^{u^i}$.
 \sk
 \noi {\sl Alternative basis}. It is convenient to use a basis of
 functions that reproduce the algebra of the $Z_n$ elements $u^i$.
 This basis is given by $y^i \equiv \sum^{n-1}_{j=0} q^{ij}
 x^{u^j}$, where $q \equiv (-1)^{2\over n}$ is the $n$-th root of
 unity. Thus $y^i y^j = y^{i+j}$, $y^0=I$. For example $y^1=y$ is
 given by
 \eq
 y=x^e+qx^u+q^2 x^{u^2}+...q^{n-1} x^{u^{n-1}} \label{newy}
 \en
Using $\sum_{j=0}^{n-1} q^{ij}=n~ \de_{i,0}$ one finds the inverse
transformation: $x^{u^i}={1\over n} \sum_{j=0}^{n-1} q^{-ij} y^j$.
Finally the $G$ action is: $\Lcal_{u^j} y^i=\Rcal_{u^j} y^i=
q^{ij} y^i$
 \sk
 \noi {\sl Conjugation classes}:
 $\{e\},\{u\},\{u^2\},...,\{u^{n-1}\}$. Unions of different
 (nontrivial) conjugation classes give rise to different calculi.
 We'll use the differential calculus corresponding to the single
 conjugation class $\{u\}$.
 \sk
 \noi {\sl Left-invariant one-forms}: $\theta^{u^i} =  \sum^{n-1}_{j=0}
 x^{u^{i+j}} dx^{u^j}$. In the one-dimensional bicovariant calculus we are
 interested in, all the $\theta^{u^i}$ are set to zero, except
  \eq
  \theta^u=  \sum^{n-1}_{j=0} x^{u^{j+1}} dx^{u^j}= \sum^{n-1}_{j=1}
  (x^{u^{j+1}}-x^u) dx^{u^j} = -\theta^e
  \en
 Thus the only independent left (and right)-invariant one-form is
 $\theta^u$.
 \sk
 \noi {\sl Inversion formula}:
 $dx^{u^i}=(x^{u^{i+1}}-x^{u^i})~\theta^u$, or in the $y$ basis:
 $dy^i= (\Rcal_{u^{-1}}y^i - y^i)~\theta^u =(q^{-i}-1)y^i~
 \theta^u$. Whereas $\theta^u$ cannot be expressed by means of a
 single differential $dx^{u^i}$, it can be expressed in terms of
 a single $dy^i$: $\theta^u = {1\over q^{-i}-1}y^{n-i} dy^i$.
 Therefore any $dy^i$ can be used as basis for
 one-forms. We'll choose for simplicity $dy$.
 \sk
 \noi {\sl Independent differential in the y basis}:
 \eq
 dy=(q^{-1}-1)y~\theta^u=
 \sum^{n-1}_{j=0}(q^{j-1}-q^j)x^{u^j}\theta^u;~~~\theta^u =
 {1\over q^{-1}-1} y^{n-1} dy=\sum^{n-1}_{j=0}{1\over q^{j-1}-q^j}
 x^{u^j}dy
 \en
 Thus any one-form can be expanded on the $\theta^u$ (vielbein)
 basis or on the ``coordinate" basis $dy$. The transition from one
 basis to the other is given by the $1 \times 1$ vielbein components:
 \eq
 \theta^u_y = {1\over q^{-1}-1} y^{n-1} \equiv J
 \en
 \noi {\sl Commutations}:
 \eq
 f\theta^u=\theta^u \Rcal_u f,~~ fdy=dy~ \Rcal_u f ~~\Longrightarrow~~ x^{u^i}
 \theta^u=\theta^u x^{u^{i-1}},~~ x^{u^i}dy=dy~x^{u^{i-1}}
 \en
 \noi {\sl Partial derivatives}:
 \eq
 df=(t_uf)\theta^u=(\part_y f)dy
 \en
 where $t_u$ is the tangent vector $t_u \equiv \Rcal_{u^{-1}} -id$ and
 $\part_y$ is the curved partial derivative:
 \eq
 \part_y f = (\Rcal_{u^{-1}}f -f)J
 \en
 \noi {\sl Exterior products}
 \eq
 \theta^u \we ... \we \theta^u=0
 \en
 and similar for products of $dy$.
 \sk
 \noi {\sl Cartan-Maurer equation}:
 \eq
 d\theta^u=0
 \en
 Torsion and curvature vanish for any connection $\om^u_{~u}=c~
 \theta^u$.
 \sk
 \noi {\sl Integration}: the volume form is $\theta^u$, and the
 integral on any 1-form $\rho$ is:
 \eq
 \int \rho = \int \rho_u \theta^u = \int \rho_u~ vol= \sum_{g \in
 Z_n} \rho_u(g)
 \en
 Integration by parts holds since:
 \eq
 \int df = \int (t_uf) \theta^u = \sum_{g \in Z_n}
 (\Rcal_{u^{-1}}f-f)(g)=0
 \en
 \noi {\sl Conjugation}
 \sk
 Using (\ref{conj1}) or (\ref{conj2}):
 \eq
 (x^{u^i})^*=x^{u^i},~~~(x^{u^i})^\star=x^{u^{n-i}},
 ~~~y^*=y^{-1},~~~y^\star=y
 \en
 \sk
 \noi {\bf Calculus on $(Z_n)^N$}
 \sk
 The basis functions are just tensor products of the basis
 functions of the single $Z_n$ factors: $x^{u^{i_1}} \otimes
 ...\otimes x^{u^{i_n}}$ etc. We use the bicovariant calculus
 corresponding to the union of the N conjugation classes
 $\{u \otimes e \otimes ...\otimes e \}$, ...
 $\{e \otimes e \otimes ...\otimes u \}$. Then the N left-invariant
 one-forms are: $\theta^{u\otimes e \otimes ...\otimes e}$, ...
 $\theta^{e \otimes e \otimes ...\otimes u}$, and
  $\theta^{e \otimes e \otimes ...\otimes e}$ is minus their sum.
 For short we label the N independent  vielbeins $\theta$ as:
 $\theta^1$, $\theta^2$, ...$\theta^N$, the corresponding
 tangent vectors as $t_1$, $t_2$, ...$t_N$, etc. Moreover the
 N special group elements $(u \otimes e \otimes ...\otimes e)$, ...
 $(e \otimes e \otimes ...\otimes u)$ are likewise denoted $u_1,
 ...u_N$, and the $N$ special $y$ coordinate functions follow the same notation:
 $y^1 = (y\otimes id ...\otimes id)$ etc.  Thus, for example:
 \eq
 df=(\Rcal_{u_1^{-1}}f-f)\theta^1 + ... +
 (\Rcal_{u_N^{-1}}f-f)\theta^N = \sum_i (t_i f) \theta^i
 \en
 and
 \eq
 dy^i=(q^{-1}-1)y^i \theta^i,~~~\theta^i = {1 \over q^{-1}-1}
 (y^i)^{n-1} dy^i
 \en
 the vielbein components being therefore diagonal $\theta^i_j=
  {1 \over q^{-1}-1} (y^i)^{n-1} \de^i_j$.

 Commutations between one-forms and functions are simply given by
  \eq
 f\theta^i=\theta^i \Rcal_{u_i} f,~~ fdy^i=dy^i~ \Rcal_{u_i} f
 \label{fdy}
 \en
Curved partial derivatives:
 \eq
 \part_{y^i} f = (\Rcal_{u^{-1}_i}f -f)J_i,~~~J_i \equiv
 {1\over q^{-1}-1} (y^i)^{n-1}
 \en
The exterior products are antisymmetric (as for any abelian group,
see the defining formula (\ref{exprod}). The Cartan-Maurer
equations still read $d\theta^i=0$. The volume form can be taken
to be
 \eq
 vol=\theta^1 \we ...\we \theta^N
 \en
 and the $\epsilon$ tensor in this case coincides with the usual
 Levi-Civita alternating tensor. Integration by parts holds
 because of $d\theta^i=0$.

 \section{Yang-Mills theory on finite $G$}
\sk
  Notations: $\Gcal$ indicates the gauge group, which we take to be
  a unitary Lie group.
  \sk
 \noi {\bf Gauge potential}
 \sk
 The dynamical field of finite $G$ gauge theory is a matrix-valued
  one-form $A(x)=A_i(x) \theta^i$. As in the usual case, $\Gcal$-gauge
  transformations are defined as
  \eq
  A'=-(dG)G^{-1}+GAG^{-1} \label{gaugeA}
  \en
  where $G(x)$ is a $\Gcal$ unitary group element ($G^\dagger=G^{-1}$)
  in some irrep, depending on
  the coordinates $x$ of the finite $G$ group manifold.
   The $\dagger$ conjugation acts as $*$ on the $x$ coordinates.
    In components:
  \eq
  A'_h=-(\part_h G) \Rcal_{h^{-1}}
  G^{-1}+GA_h \Rcal_{h^{-1}}G^{-1}
  \en
  Notice that $\part_hG$ is a finite difference of group elements,
  and therefore $A_h$ must belong to the $\Gcal$ {\sl group algebra},
  rather than to the $\Gcal$ Lie algebra.
  \sk
  \noi {\bf Hermitian conjugation}
  \sk
  We define hermitian conjugation on matrix valued one forms
  $A$ as follows:
  \eq
  A^\dagger = (A_h \theta^h)^\dagger \equiv (\theta^h)^*
  A_h^\dagger=-\theta^{h^{-1}} A^\dagger_h=-\theta^h
  A^{\dagger}_{h^{-1}}
  \en
  where $\dagger$ acts as hermitian conjugation on the matrix
  structure of $A_h$ and as $*$ conjugation on the $Fun(G)$
  entries of the matrix.
  \sk
  \noi {\bf Matter fields and covariant derivative}
  \sk
  Matter fields $\psi$ transform in an irrep of $\Gcal$:
  \eq
  \psi'=G\psi,~~~(\psi^\dagger)'=\psi^\dagger G^\dagger =
  \psi^\dagger G^{-1}
  \en
  and their covariant derivative, defined by
  \eq
  D\psi= d\psi +A \psi, ~~~D\psi^\dagger= d\psi^\dagger -
  \psi^\dagger A
  \en
  transforms as it should: $(D\psi)'=G (D\psi),~(D\psi^{\dagger})'=
 (D\psi^{\dagger})G^{-1}$. Requiring compatibility of hermitian
 conjugation with the covariant derivative , i.e.
 $(D\psi)^{\dagger}=D \psi^{\dagger}$, implies:
  \eq
  A^{\dagger}=-A
  \en
  that is, $A$ must be an antihermitian connection. In components
  this means:
  \eq
  A^\dagger_h=\Rcal_{h^{-1}} A_{h^{-1}}
  \en
  \sk
  \sk
 \noi {\bf Field strength}
 \sk
 The field strength $F$ is formally defined as usual:
 \eqa
 & & F=dA+A\we A=d(A_k\theta^k)+A_h\theta^h \we A_k
 \theta^k=\nonumber \\
 & & =(\part_h A_k) \theta^h\we\theta^k + A_k d\theta^k + A_h
 (\Rcal_{h^{-1}} A_k) \theta^h \we \theta^k = \nonumber\\
 & &~~~= [\part_h A_k - A_j C^j_{~hk} + A_h (\Rcal_{h^{-1}} A_k)]
 \theta^h \we \theta^k
 \ena
 and varies under gauge transformations (\ref{gaugeA}) as
 \eq
 F'=GFG^{-1}
 \en
 or in components:
 \eq
 F'_{hk}=GF_{hk} \Rcal_{h^{-1}k^{-1}} G^{-1},~~~(F^\dagger_{hk})'=
(\Rcal_{h^{-1}k^{-1}} G)F^\dagger_{hk} G^{-1}
 \en
 \noi {\bf Action}
 \sk
 Due to the above gauge variations of $F$, the following action
 is invariant under gauge transformations:
 \eq
 A_{YM}=\sum_G Tr (F_{hk} F^\dagger_{hk}) \label{YMaction}
 \en
 \noi {\bf Link variables}
 \sk
 Introducing the link field $U_h(x)$:
 \eq
 U_h(x) \equiv \unop + A_h(x)
 \en
 transforming as
 \eq
 U'_h=GU_h\Rcal_{h^{-1}} G^{-1}
 \en
 the $F$ components take the form
 \eq
 F_{hk}=U_h \Rcal_{h^{-1}} U_k - U_k \Rcal_{k^{-1}} U_{khk^{-1}}
 \label{FasU}
 \en
 Requiring $U_h$ to be unitary ($U^\dagger_h=U^{-1}_h$), and
 substituting (\ref{FasU}) in the action (\ref{YMaction}) leads to
 a suggestive result
 \eq
 A_{YM} = - \sum_G Tr[U_h (\Rcal_{h^{-1}} U_k) (\Rcal_{k^{-1}}
 U^{-1}_{khk^{-1}}) U^{-1}_k + herm.~ conjugate] \label{YMactionU}
 \en
 (a constant term $2 \sum_G Tr \unop$ has been dropped).
 When the finite group $G$ is abelian, the action
 (\ref{YMactionU}) reduces to the Wilson action. In particular
 this happens for $G= Z_N \times Z_N \times Z_N \times Z_N$, a
 result already obtained in ref. \cite{DMgauge}.
 \sk
 \noi {\bf Fermion coupling}
 \sk
 We can add a Dirac term for fermions $\psi$:
 \eq
 A_{Dirac} = \sum_G \psi^\dagger \ga^0 \ga^h D_h \psi
 \en
 invariant under global Lorentz transformations $SO(dimG')$ and
 local $\Gcal$ gauge transformations.

 \section{Invariant $G$-gravity actions}

 We have now all the ingredients necessary for the construction of
 gravity actions on finite group spaces, invariant under the analogues of diffeomorphisms and
 local Lorentz rotations.

 What we aim for is a dynamical theory of vielbein fields whose
 ``vacuum" solution describes the finite $G$ manifold. Then the dynamical
 vielbeins $V^a$ are not left-invariant any more, being a deformation of
 the $\theta$ one-forms:
 \eq
 V^a = \sumongp V^a_g (x) \theta^g
 \en
 The vielbein components $V^a_g$  along the rigid basis $\theta$ are
 assumed to be invertible $x$-dependent matrices. The inverse we
 denote as usual by $V^g_a$.

 In addition we also consider the spin connection 1-form
 $\om^{g_1}_{~g_2}$ as an independent field (first order
 formulation). The $\om$ field equations will then determine the
 expression of $\om$ in terms of the vielbein field.

 We propose two different actions for the $G$-analogue of gravity:
 \eq
  A_G = \int R ~ \det (V^a_i) vol(G) = \sumong R ~\det(V^a_i) \label{A1}
 \en
 \noi
 and
 \eq
 A_G = \int R ~ \epsilon _{a_1...a_p}V^{a_1}_{i_1} ...V^{a_p}_{i_p}
 \theta^{i_1} \we ...\we
 \theta^{i_p} = \sumong R ~det_{\epsilon}( V^a_i) \label{A2}
 \en
 In both actions the scalar $R$
 is the finite group analogue of the Gaussian curvature:
 \eq
 R \equiv V^h_a (\Rcal_{h^{-1}k^{-1}}V^k_b )
 R^{ab}_{~~~hk},~~~R^{ab}_{~~~hk} \equiv \ga^{bc} R^a_{~c~hk}
 \label{gauss}
 \en
 where $\ga^{bc} = \de^{bc}$, and the curvature components
 on the rigid basis $\theta$ are
 defined by $R^a_{~b} = R^a_{~b~hk} \theta^h \otimes \theta^k$:
 \eq
 R^a_{~b~hk} = \part_h \om^{a}_{~b~k} - \part_k
 \om^{a}_{~b~khk^{-1}}-\om^{a}_{~b~i}\Cb^i_{~h,k}+\om^{a}_{~c~h}
 (\Rcal_{h^{-1}} \om^{c}_{~b~k})-
 \om^{a}_{~c~k}(\Rcal_{k^{-1}} \om^{c}_{~b~khk^{-1}} )
 \en
 where the constants $\Cb$ are given by the $G$-antisymmetrization of
 the $C$ constants (cf. eq. (\ref{Cconst})): $\Cb^i_{~h,k} \equiv
 C^i_{~h,k} - C^i_{k,khk^{-1}}$.

 The determinant in (\ref{A1}) is the usual determinant of the
 $m \times m$ matrix $V^a_i$, while the ``determinant" in (\ref{A2})
 is computed via the $\epsilon$
 tensor of eq. (\ref{vol}), i.e. $det_{\epsilon}(V^a_i) = \epsilon^{i_1...i_p} \epsilon_{a_1...a_p}
 V^{a_1}_{i_1}...V^{a_p}_{i_p}$,  $p$ being the order of the volume form.
 \sk
 \noi {\bf Invariances of $A_G$}
 \sk
 Both actions are invariant under the local field transformations:
 \eqa
 & &(V^{b'}_{h})'=a^{b'}_{~b} V^b_h \label{aV} \\
 & &(\om^{b'}_{~c'})'=a^{b'}_{~b} \om^{b}_{~c} (a^{-1})^{c}_{~c'}
 + a^{b'}_{~c} d(a^{-1})^c_{~c'} \label{aom}
 \ena
 where $a$ is an $x$-dependent matrix that rotates the metric and the
 $\epsilon$ tensor into themselves. For the first action $a$
 simply belongs to $SO(m)$, while for the second it belongs to
 a subgroup of $O(m)$ that preserves the $\epsilon$ tensor of eq. (\ref{vol}).
 These are then the local tangent invariances of the two actions.
  Note that the two actions coincide in
 the case of $G =(Z_n)^N$, the $\epsilon$ tensor in (\ref{vol})
 becoming just the usual alternating tensor.

{\sl Proof:}
 under the above transformation  the curvature components vary
 according to eq. (\ref{Rtransf}):
 \eq
 (R^{b'c'}_{~~~hk})'= a^{b'}_{~b} (\Rcal_{h^{-1}k^{-1}}
 a^{c'}_{~c}) R^{bc}_{~~~hk}
 \en
(use also (\ref{ginv})), so that the Gaussian curvature in
(\ref{gauss}) is indeed invariant. So is $\det(V^a_i)$ if $a \in
SO(m)$, and the first action $A_G$ is therefore invariant under a
local $SO(m)$ tangent group. Similarly $det_{\epsilon}(V^a_i)$ is
a scalar under the $O(m)$ transformations conserving $\epsilon$, a
subgroup of $O(m)$.
  \sk
 {\bf Note}. The two actions above correspond
 to two different definitions of
  the volume of the deformed $G$ manifold:
  \eq
 vol({\tilde G}) \equiv \int \det(V^a_i)~ vol(G)
 \en
\noi and
 \eq
 vol({\tilde G}) \equiv \int \epsilon _{a_1...a_p}V^{a_1}_{i_1} ...V^{a_p}_{i_p} \theta^{i_1} \we ...\we
 \theta^{i_p}
 \en
 with the same local symmetries as for the corresponding actions.
 In the second case, we are not using the seemingly more natural expression
 $\int \epsilon _{a_1...a_m}V^{a_1} \we ...\we V^{a_m}$ since it is not
 invariant under (\ref{aV}).
 \sk
 Both $A_G$ are also invariant under infinitesimal diffeomorphisms
  generated by the Lie derivative $\ell_V$ along an arbitrary tangent vector $V$,
  if integration by parts holds. Indeed the variation of any
  action $A$ reads:
  \eq
  \de A = \int \ell_V ( p\mbox{-form}) = \int [di_V (p\mbox{-form}) + i_V d
  (p\mbox{-form})]=0
\en
since $d(p\mbox{-form})=0$ and $\int d(p-1\mbox{~-form})=0$.
 \sk
  The invariance under the finite coordinate transformations
  (\ref{gct}) is somewhat trivial, since there are no ``world" indices
  in the definition of $A_G$ that transform under it. It would be
  possible, in principle, to introduce world indices by expressing
  the deformed vielbein $V^a_i$ as
  \eq
  V^a_i (x) \equiv V^a_\al (x) \theta^\al_i (x)
  \en
  \noi where $\theta^\al_i$ is a given matrix (function of $x$) whose
  inverse is defined by $\theta^i \equiv \theta^i_\al dx^{\al}$.
  Then the dynamical fields are the vielbein components
  $V^a_{\al}$ transforming under both the local tangent group and
  the finite diffeomorphisms. However this would clearly be
  artificial, since anyhow in (\ref{A1}) and (\ref{A2}) we have to refer
  explicitly to a preferred frame of reference, spanned by
  the $\theta^i$. The reason is that only in terms of
  this frame we are able to express the commutation rules as in
  eq. (\ref{fthetacomm}). A general mixture of the $\theta$ has a
  complicated commutation rule with a generic function $f$, hardly
  suited for constructing an invariant action. If one could use
  a basis of differentials $dx$ for the various forms one could
  give meaning to finite coordinate invariance: in general this is not
  fruitful since again the differentials have no simple commutations
  with the functions. A notable exception is provided by $G=(Z_n)^N$
  (and in general by any abelian group)
   as we discuss below. In this case
  the differentials $dy$ have the same commutation rules
  with functions as the $\theta$, and we find an action
  entirely written in terms of differentials $dy$ and components
  $V^a_{\al}$ explicitly invariant under {\sl linear} general
  coordinate transformations.

  \sk
    \noi {\bf Field equations}
  \sk
  Varying the actions $A_G$ with respect to the vielbein
  $V^c_j$ yields the analogue of the Einstein equations,
  respectively
  \eq
   R^j_{~c}
  +\Rcal_{kh} (V^h_a R^{a~b}_{~~~h~k} \det V ) (\det V)^{-1} V^k_c V^j_b - R~ V^j_c=0
  \label{Einsteineq1}
  \en
  and
  \eq
  R^j_{~c}~
  det_{\epsilon} V+\Rcal_{kh} (V^h_a R^{a~b}_{~~~h~k} ~det_
  {\epsilon} V)~ V^k_c V^j_b -
  p~ R ~ \epsilon^{j~i_2...i_p} \epsilon _{c~a_2...a_p} V^{a_2}_{i_2} ...
  V^{a_p}_{i_p} =0
  \label{Einsteineq}
  \en
  where $R^j_{~c}$, the analogue of the Ricci tensor, is defined by
  \eq
  R^j_{~c}= V^h_c V^j_a (\Rcal_{h^{-1}k^{-1}} V^k_b) R^{a~b}_{~~~h~k}
  \en
  and $R=V^c_j R^j_{~c}$.

Similarly varying the actions (\ref{A1}), (\ref{A2}) with respect
to $\om^{ab}_{~~i}$ yields a system of linear equations for all
the components of the spin connection. We will write it explicitly
in the case of $(Z_n)^N$.

 \sk
 \noi {\bf Gravity action on $(Z_n)^N$ }
 \sk
 In this case we can write the action
 \eq
 A = \int R \det V~ d^Ny \label{A3}
 \en
 where now the curvature scalar is given by
  \eq
 R \equiv V^\al_a (\Rcal_{\al^{-1}\be^{-1}}V^\be_b )
 R^{ab}_{~~~\al\be},~~~R^{ab}_{~~~\al\be} \equiv \ga^{bc} R^a_{~c~\al\be}
 \label{gaussZ}
 \en
 and the curvature components are defined in the usual way as $R^{ab} \equiv
 R^{ab}_{~~\al\be} ~dy^{\al} \we dy^{\be}$:
 \eq
 R^{ab}_{~~\al\be}=\part_\al \om^{ab}_{~~\be}+\om^{ac}_{~~\al}
 (\Rcal_{\al^{-1}} \om_{c~\be}^{~b}) -(\al \leftrightarrow \be)
 \en

  The determinant in
 (\ref{A3}) is the usual determinant of the vielbein field
 $V^a_{\al}$, and the volume element is the usual $d^N y \equiv
 \epsilon_{\al_1...\al_N} dy^{\al_1} \we ... \we dy^{\al_N}$.
 Since differentials and $\theta$ have the same commutations with functions
 (see eq. (\ref{fdy}), the action is again invariant under the
 local $SO(N)$ transformations $(V^{b'}_{\be})'=a^{b'}_{~b} V^b_\be$ and
 (\ref{aom}). The action $A$ is also invariant under the finite
 coordinate transformations:
 \eq
 y'^{\al} = y'^{\al}_{0}I + y'^{\al}_{\be 1} y^\be \label{gctlin}
 \en
 \noi with $y'^{\al}_{0}, y'^{\al}_{\be 1}$ real constants, so that
  the transition function $y'^{\al}(y)$ is linear in the
 old coordinates $y^{\be}$. The curved derivatives  $\part_{\be} y'^{\al}$
 are then constant, and commute therefore with all the differentials $dy$.
 The volume $d^Ny$ transforms under (\ref{gctlin}) with the
 determinant of the curved derivative matrix (jacobian) while $\det V$
 transforms with the inverse jacobian, so that $\det V d^Ny$ is
 invariant under (\ref{gctlin}). The field equations are as in eq.
 (\ref{Einsteineq1}), after replacing all $j,h,k$ indices with
 curved indices $\al,\be,\ga$. The ``vacuum" solution
 $R^{ab}_{~~\al\be}=0$ corresponds to the vielbein
 $V^a_{\al}=\de^a_{\al}$ which describes the rigid $(Z_n)^N$
 manifold, the discrete analogue of flat euclidean space.
 \sk
  \noi {\sl Calculating $\om$ in terms of $V$}
  \sk
  Varying the action (\ref{A3}) with respect to $\om^{ab}_{~~\be}$
   yields the analogue of the torsion equation:
  \eq
  J_\al (id - \Rcal_\al) W^{\al\be}_{ab}+(\Rcal_{\be^{-1}}
  \om_{b~~\al}^{~c}) W^{\be\al}_{ac}+ q\Rcal_\al [\om^c_{~a\al}
  W^{\al\be}_{cb}]=0
  \en
   \noi (no sum on $\be$) with
  \eq
  W^{\al\be}_{ab} \equiv [V^\al_a(\Rcal_{\al^{-1}\be^{-1}}
  V^\be_b)-(\al \leftrightarrow \be)] ~\det V
  \en
 That a solution for $\om$ exists can be verified in simple cases,
 as for example $G= Z_2 \times Z_2$. A simplifying assumption
 consists in taking $\om^{ab}_{~~\be}$ to be antisymmetric in
 $a,b$.

\section{Conclusions}

Differential calculi on discrete spaces are a powerful tool in the
formulation of field theories living in such spaces. These calculi
are in general noncommutative, and are constructed algebraically.
We can expect them to be of relevance also in understanding the
noncommutative field theories arising from string/brane theory.


\vfill\eject
\end{document}